\documentclass[twocolumn,prl]{revtex4}

\usepackage{amsfonts,amsmath,amssymb,amsthm}
\usepackage{epic,eepic}
\usepackage{graphicx}
\usepackage{subfigure}

\usepackage{setspace}

\newcommand{\lmax}{\lambda_{\rm max}}
\newcommand{\lmin}{\lambda_{\rm min}}

\newtheorem*{theorem*}{Theorem}

\newcommand{\la}{\lambda}
\newcommand{\eps}{\epsilon}
\newcommand{\aff}{a_{\rm ff}}
\newcommand{\afb}{a_{\rm fb}}

\newcommand{\bei}{\begin{itemize}}
\newcommand{\eei}{\end{itemize}}
\newcommand{\beq}{\begin{equation}}
\newcommand{\eeq}{\end{equation}}
\newcommand{\beqr}{\begin{eqnarray}}
\newcommand{\eeqr}{\end{eqnarray}}
\newcommand{\beqrn}{\begin{eqnarray*}}
\newcommand{\eeqrn}{\end{eqnarray*}}
\newcommand{\brr}{\begin{array}}
\newcommand{\err}{\end{array}}
\newcommand{\bef}{\begin{figure}}
\newcommand{\eef}{\end{figure}}

\begin{document}

\title{Reliable and unreliable dynamics in driven coupled oscillators}


\author{Kevin K. Lin}
\author{Eric Shea-Brown}
\author{Lai-Sang Young}

\affiliation{Courant Institute of Mathematical Sciences and
Center for Neural Science \\New York University}

\date{August 8, 2006}

\begin{abstract}

This letter concerns the \textit{reliability} of coupled
oscillator networks in response to fluctuating inputs.
Reliability means that (following a transient) an input elicits
identical responses upon repeated presentations, regardless of
the system's initial condition.  Here, we analyze this property
for {\it two coupled oscillators}, demonstrating that oscillator
networks exhibit both reliable and unreliable dynamics for broad
ranges of coupling strengths.  We further argue that unreliable
dynamics are characterized by strange attractors with random SRB
measures, implying that though unreliable, the responses lie on
low-dimensional sets.  Finally, we show that 1:1 phase locking
in the zero-input system corresponds to high susceptibility for
unreliable responses.  A geometric explanation is proposed.

\end{abstract}

\maketitle


For a dynamical system, the question of reliability can be
formulated as follows: if an external stimulus is applied
multiple times, will it elicit essentially the same response
each time {\it independent of the state of the system} when the
input is received? The answer is fundamental to the ability of
the system to encode stimuli in a repeatable way. It also
determines if an ensemble of such systems would synchronize
if they receive the signal as a common input. Applications range
from biological pacemakers to laser arrays to neurons processing
incoming stimuli~\cite{general_books}.

The reliability of isolated oscillators has been explored
extensively in both experiment and theory.  For example, single
neurons respond reliably to fixed-current signals in laboratory
experiments {\cite{mainen-etc}}.  Theoretical studies
{\cite{reliable,reliable_and_amp_effects}} have shown that
reliability is typical for phase oscillator models, {\em i.e.}
ODEs on the circle.  Unreliable dynamics, characterized by an
apparently chaotic response to the stimulus, have been shown to
occur if the state space of the oscillator has more than one
dimension and the stimulus is sufficiently
strong~\cite{unreliable}.

The reliability of {\em networks} of coupled oscillators is far
less well understood.  Here, we take a first step toward
understanding how input stimulus and network architecture
interact to determine reliability.  For concreteness, we focus
on a 2-oscillator pulse-coupled network.

\textbf{Our main findings are:} {\bf (1)} Single oscillators
that are reliable in isolation can become highly unreliable when
coupled, depending on input amplitude and coupling strengths.
{\bf (2)} Strange attractors with random SRB measures are a
signature of unreliability. The singularity of these measures
implies that even in unreliable dynamics, phase relations are
highly structured.  {\bf (3)} Parameters at which the two-cell
network is especially susceptible to producing unreliable
dynamics coincide with the onset of 1:1 phase locking for the
system with zero input.  Geometric explanations for some of the
observed phenomena are proposed.


\medskip
\centerline{\bf \small I. THE TWO-CELL MODEL}

\medskip

\begin{figure*}
\resizebox{3.2in}{!}{\includegraphics[bb=0in 0in 9.6in 3in]{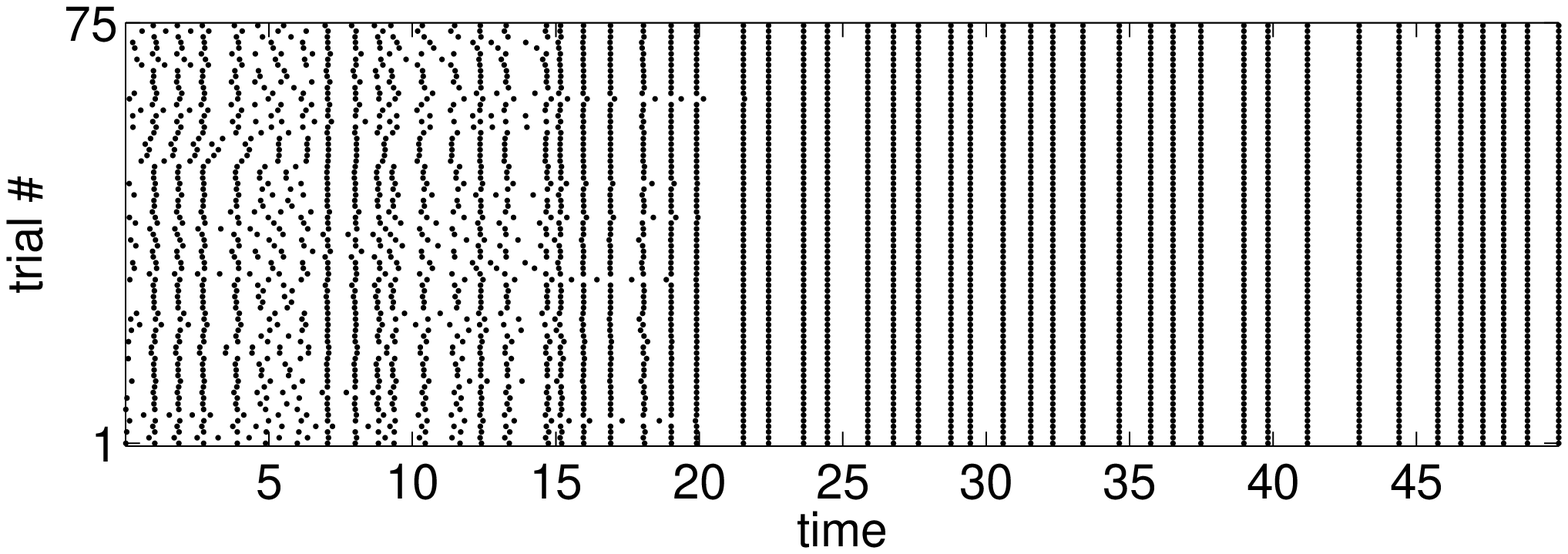}}\hspace*{6pt}%
\resizebox{3.2in}{!}{\includegraphics[bb=0in 0in 9.6in 3in]{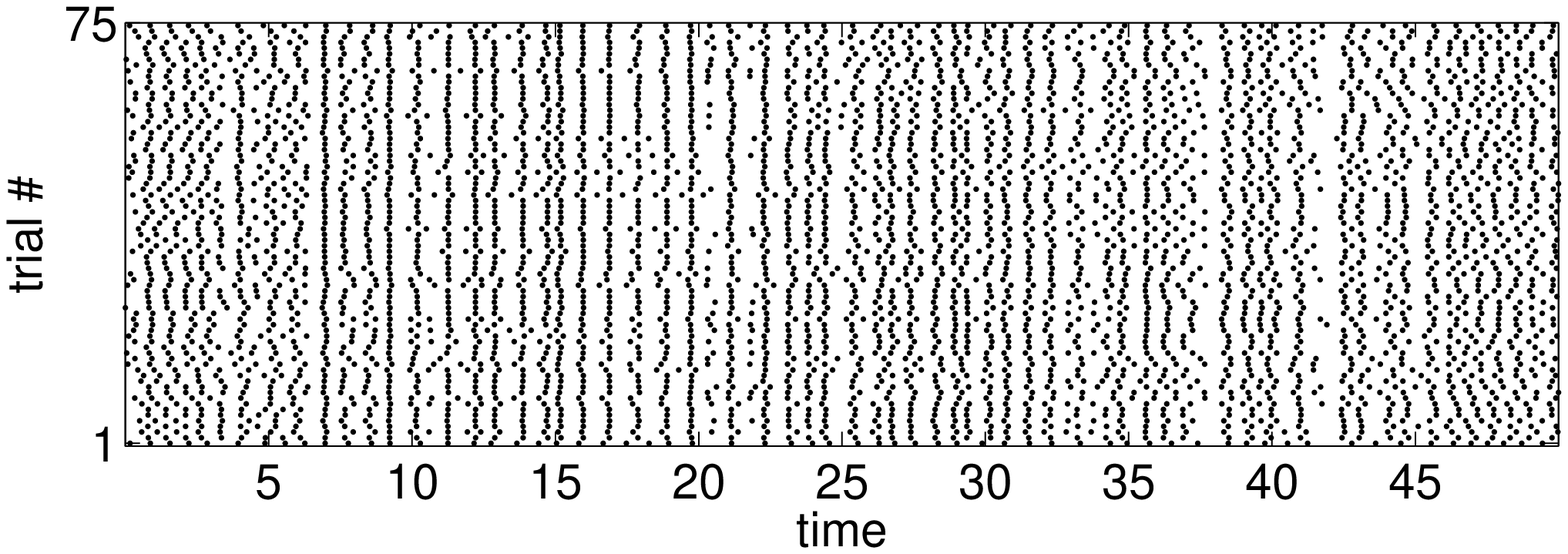}}
\caption{Raster plots.  In each experiment, the results of 75 trials are shown, and dots are placed at spike times of cell 1 (with each trial corresponding to one row).  {\bf
Left:} feedforward network ($\aff=1$, $\afb=0$), exhibiting
identical spike times or reliability.  {\bf Right:} Positive
feedback ($\aff=1$, $\afb=1.5$) with unreliable responses.}
\label{f.rasters}
\end{figure*}

We consider in this letter a 2-cell network of oscillators
with the following structure:
\begin{center}
\setlength{\unitlength}{0.0004in}
\begingroup\makeatletter\ifx\SetFigFont\undefined%
\gdef\SetFigFont#1#2#3#4#5{%
  \reset@font\fontsize{#1}{#2pt}%
  \fontfamily{#3}\fontseries{#4}\fontshape{#5}%
  \selectfont}%
\fi\endgroup%
{\renewcommand{\dashlinestretch}{30}
\begin{picture}(4533,1113)(0,-10)
\put(2625,849){\makebox(0,0)[lb]{{\SetFigFont{10}{14.4}{\rmdefault}{\mddefault}{\updefault}$\aff$}}}
\put(3916.875,502.181){\arc{808.949}{5.2529}{7.4711}}
\blacken\path(4236.560,795.575)(4125.000,849.000)(4198.294,749.361)(4236.560,795.575)
\put(1575,549){\ellipse{1082}{1082}}
\put(3900,549){\ellipse{1082}{1082}}
\thicklines
\path(2025,549)(2175,549)
\path(4350,549)(4500,549)
\thinlines
\path(3225,399)(2325,399)
\blacken\path(2445.000,429.000)(2325.000,399.000)(2445.000,369.000)(2445.000,429.000)
\path(2325,699)(3225,699)
\blacken\path(3105.000,669.000)(3225.000,699.000)(3105.000,729.000)(3105.000,669.000)
\put(3825,474){\makebox(0,0)[lb]{{\SetFigFont{10}{14.4}{\rmdefault}{\mddefault}{\updefault}$\theta_2$}}}
\put(1500,474){\makebox(0,0)[lb]{{\SetFigFont{10}{14.4}{\rmdefault}{\mddefault}{\updefault}$\theta_1$}}}
\put(-600,400){\makebox(0,0)[lb]{{\SetFigFont{10}{14.4}{\rmdefault}{\mddefault}{\updefault}$\eps I(t)$}}}
\put(200,400){\makebox(0,0)[lb]{{\SetFigFont{10}{14.4}{\rmdefault}{\mddefault}{\updefault}{\huge
    $\rightsquigarrow$}}}}
\put(2625,99){\makebox(0,0)[lb]{{\SetFigFont{10}{14.4}{\rmdefault}{\mddefault}{\updefault}$\afb$}}}
\put(1573.875,549.181){\arc{808.949}{5.2529}{7.4711}}
\blacken\path(1893.560,842.575)(1782.000,896.000)(1855.294,796.361)(1893.560,842.575)
\end{picture}
}
\end{center}

To illustrate our ideas, we use a standard phase model which
characterizes the dynamics of limit cycle oscillators with smooth,
pulsatile interactions \cite{model}:
\begin{align}
\label{e.tht} \dot\theta_1 &= \omega_1 + a_{\rm fb} g(\theta_2)
z(\theta_1) + \epsilon z(\theta_1) I(t)  \\
\dot\theta_2 &= \omega_2 + a_{\rm ff} g(\theta_1)
z(\theta_2),\nonumber
\end{align}
The state of each oscillator is described by a phase, i.e., an
angular variable $\theta_i \in {\mathbb S}^1 \equiv {\mathbb
R}/{\mathbb Z}, i=1,2$.  The constants $\omega_1$ and $\omega_2$
are the cells' intrinsic frequencies, $I(t)$ is the external
stimulus and $\eps$ its amplitude.  In this letter $I(t)$ is
taken to be white noise.  The coupling is via a ``bump
function'' $g$ which vanishes outside of an interval $(-b,b)$;
we have taken $b$ to be $\approx 0.05$.  On $(-b,b)$, $g$ is
smooth, it is $\geq 0$ and satisfies $\int_{-b}^b{g(\theta)\
d\theta} = 1$. The meaning of $g$ is as follows: We say the
$i$th oscillator ``spikes'' when $\theta_i(t)=0$.  Around the
time that an oscillator spikes, it emits a pulse which modifies
the other oscillator.  The strength of the feedforward (ff)
coupling is given by $a_{\rm ff}$ (``forward'' refers to the
direction of stimulus propagation), and likewise for $\afb$, the
feedback connection. Finally, the response of an oscillator to
coupling and stimulus is modulated by its phase response curve
$z$~\cite{PRC,EI}. We take $z(\theta)
=\frac{1}{2\pi}(1-\cos(2\pi\theta))$, characterizing oscillators
near saddle-node bifurcations on limit cycles (as for many
neuron models~{\cite{EI}}).

The simplest way to investigate the reliability of a system is
to carry out many simulations, using a different initial
condition each time but driving the system with {\it the same}
stimulus $\eps I(t)$, and to record spike times in raster
plots. Fig. 1 contains the results for cell 1 for two distinct
sets of connection strengths. On the left is the
\textit{feedforward case}, with $\aff=1$ and $\afb=0$: After a
brief transient, every trial ({\em i.e.}  every initial state)
evolves to produce identical spike times.  This is a signature
of reliability. With $\afb=0$, the presence of cell 2 is
irrelevant; our results are thus consistent with
~\cite{reliable,reliable_and_amp_effects}. On the right is the
\textit{positive feedback case}, where $\aff=1$ and $\afb=1.5$:
The plot shows that even though some spiking events are shared,
there is no convergence to common spike times.



\medskip
\centerline{\bf \small II. RANDOM ATTRACTORS}

\medskip

With $I(t)$ taken to be realizations of white noise, we treat
(\ref{e.tht}) as a stochastic differential equation (SDE), and
its solutions as {\it random dynamical systems} (RDS).  We
review below some relevant mathematical facts and discuss their
interpretations.

{\bf Review of RDS theory \cite{RDS_review}:} Consider the SDE
\begin{equation}
\label{eq:sde}
dx_t = a(x_t)\ dt + b(x_t)\ dB_t \;,
\end{equation}
where $B_t$ is a standard Brownian motion. We assume that the
steady-state statistics of $x_t$ are governed by a stationary
probability measure $\mu$ which has a density ($\mu$ can be
found by solving the Fokker-Planck equation).  By a very general
theorem, the solution of (\ref{eq:sde}) has a representation as
an RDS: Associated with almost every sample Brownian path
$\omega$ is a family of diffeomorphisms (i.e. smooth invertible
transformations) of the phase space $\{F_{s,t;\omega}, s<t\}$
with the property that for every point $x$ in the phase space,
if $x_s=x$, then $F_{s,t;\omega}(x)=x_t$ gives the solution to
the SDE at time $t$.

Thinking of the process as starting from $t=-\infty$, we obtain a
family of {\it sample measures} $\{\mu_\omega\}$ which are
conditional measures of $\mu$ given the history of the Brownian path
up to time $0$. That is, $\mu_\omega$ describes what one sees at
$t=0$ given that the system has experienced the perturbations
defined by the realization of Brownian motion $\omega$ for $t <0$.
This property can also be expressed by
\begin{equation}
(F_{-t,0;\omega})_*\mu \to \mu_\omega  \quad {\rm as}
\quad t \to \infty
\label{muomega}
\end{equation}
where $(F_{-t,0;\omega})_*\mu$ is the measure obtained by
transporting $\mu$ forward by $F_{-t,0;\omega}$. Finally,
the family $\{\mu_\omega\}$ is invariant in the sense that
$(F_{0,t;\omega})_*(\mu_\omega) = \mu_{\sigma_t(\omega)}$
where $\sigma_t (\omega)$ is the
time-shift of the sample path $\omega$ by $t$.

{\bf Interpretation:} For us, $x_t=(\theta_1(t), \theta_2(t))$,
our phase space is ${\mathbb S}^1 \times {\mathbb S}^1$, and
each $\omega$ corresponds to a stimulus $I(t), t \in (-\infty,
\infty)$.  If our initial distribution is given by a probability
density $\rho$ and we apply the stimulus $\eps I$, then the
distribution at time $t$ is $(F_{0,t;\omega})_*\rho$. For $t$
sufficiently large, one expects in most situations that
$(F_{0,t;\omega})_*\rho$ is very close to
$(F_{0,t;\omega})_*\mu$, which by (\ref{muomega}) is essentially
given by $\mu_{\sigma_t(\omega)}$.  The time-shift by $t$ of
$\omega$ is necessary because by definition, $\mu_\omega$ gives
the conditional distribution of $\mu$ at time $0$.

Fig. 2 shows some snapshots of $(F_{0,t;\omega})_*\rho$, which,
in the latter panels (where $t\gg 1$), approximate
$\mu_{\sigma_t(\omega)}$.  The initial density $\rho$ (prior to
the presentation of $\eps I$) is the steady state of the network
under weak (amplitude 0.01) white noise~\cite{initial_footnote}.
As in Fig. 1, results for two distinct cases are shown,
feedforward in the top row and positive feedback in the bottom
row.

\begin{figure*}
\resizebox{1in}{1in}{\includegraphics*[bb=0 0 107 125]{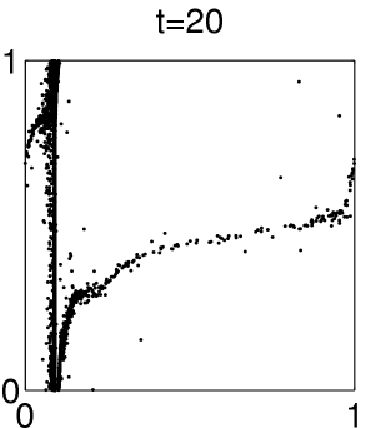}}%
\resizebox{1in}{1in}{\includegraphics*[bb=0 0 107 125]{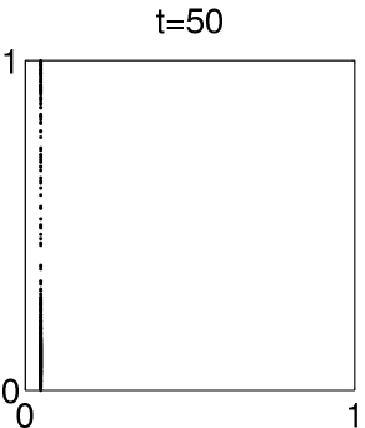}}%
\hspace*{6pt}%
\resizebox{1in}{1in}{\includegraphics*[bb=0 0 108 126]{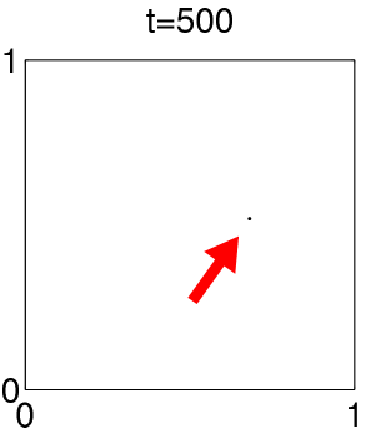}}%
\resizebox{1in}{1in}{\includegraphics*[bb=0 0 108 126]{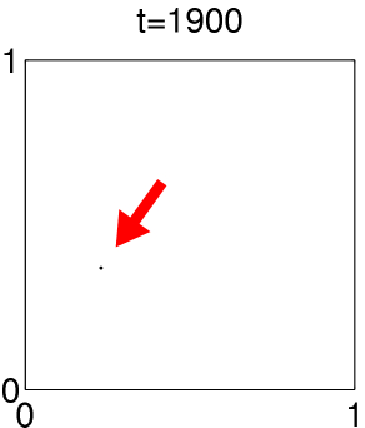}}%
\hspace*{-6pt}%
\resizebox{3in}{1in}{\includegraphics*[bb=0in 0in 6in 2in]{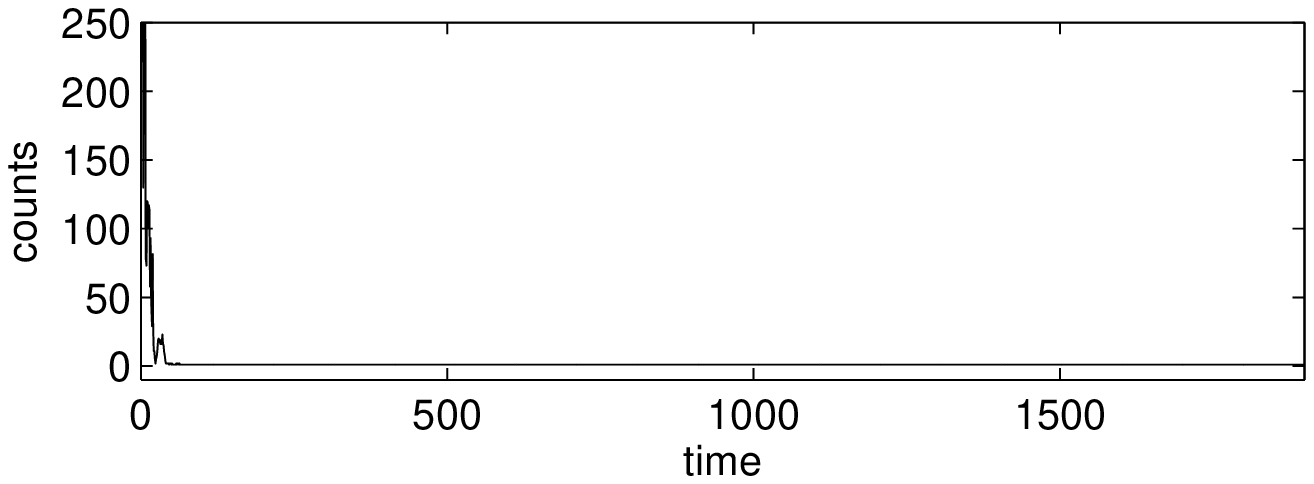}}\\
\resizebox{1in}{1in}{\includegraphics*[bb=0 0 107 125]{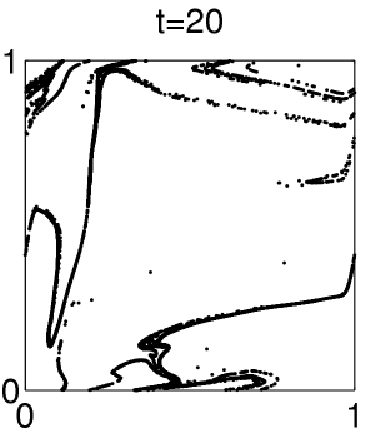}}%
\resizebox{1in}{1in}{\includegraphics*[bb=0 0 107 125]{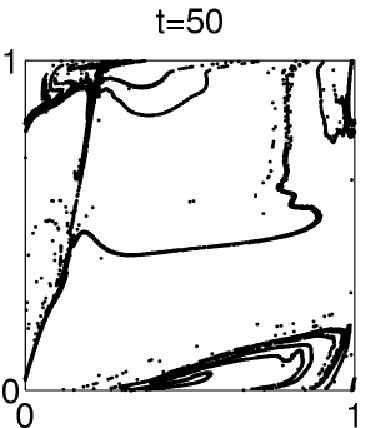}}%
\resizebox{1in}{1in}{\includegraphics*[bb=0 0 107 125]{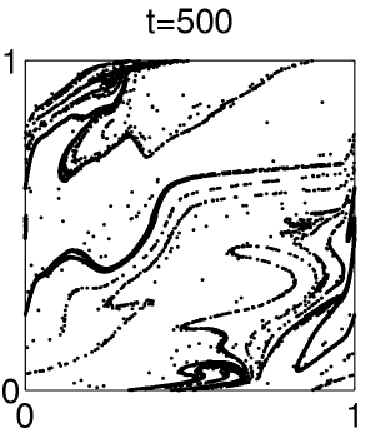}}%
\resizebox{1in}{1in}{\includegraphics*[bb=0 0 107 125]{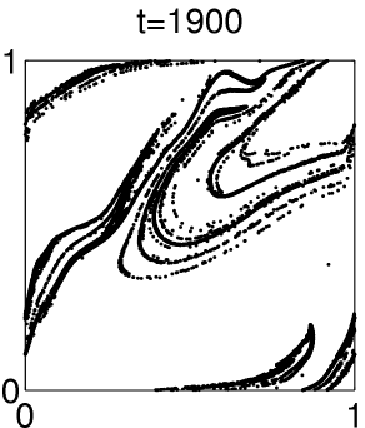}}%
\resizebox{3in}{1in}{\includegraphics*[bb=0in 0in 6in 2in]{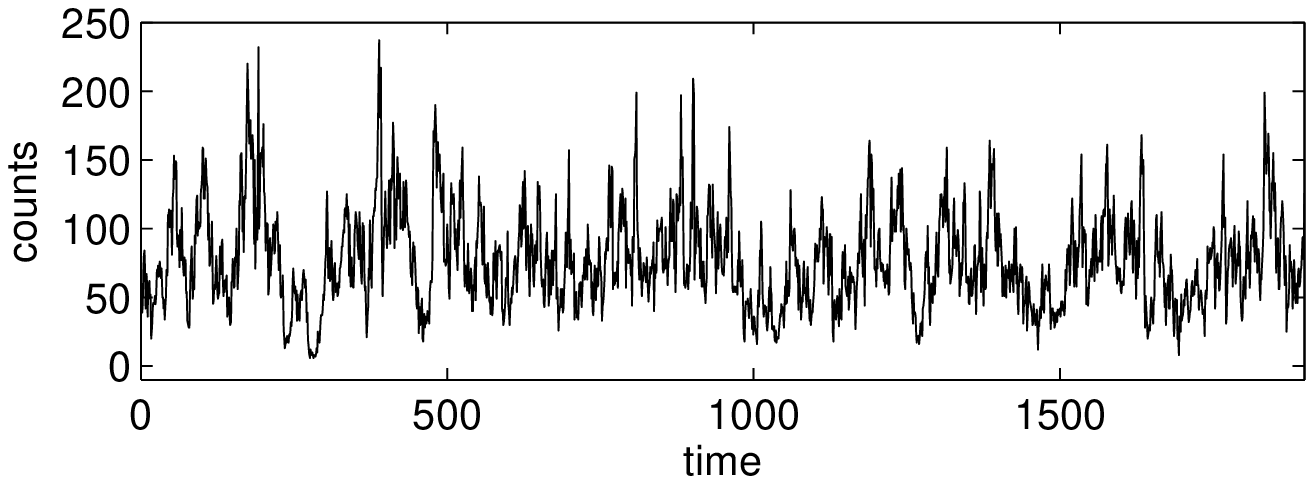}}\\
\caption{States of $50,000$ initial conditions evolved for 1900
units of time. Snapshots at various times of sample measures in
response to a single realization of the stimulus are
shown. (Top) ~\textit{Feedforward network}: the states converge
to a random fixed point.  (Bottom) ~\textit{Feedforward
network}: the states converge to random strange attractor. (Far
right) Number of grid elements in a $30 \times 30$ grid needed
to cover $90\%$ of the trajectories are plotted as a function of
time; it shows in particular that pictures of random SRB
measures with their characteristic geometry, though fluctuating
in time, are persistent.}
\label{f.snapshot}
\end{figure*}

{\bf Two relevant results:} Lyapunov exponents (LE) measure the
exponential rates of separation of nearby trajectories in a
dynamical system {\cite{er}}.  The value of the largest LE, $\lambda_{\max}$, contains a great deal of information
about the system:  A positive $\lmax$ for a large set of
trajectories is synonymous with chaos, while the presence of
stable equilibria is characterized by $\lmax<0$.  As in
deterministic systems, LE for RDS are well
defined, and they are {\it nonrandom}, i.e. they do not depend
on $\omega$.

\smallskip
\noindent THEOREMS. {\it The setting is as above;
$\mu$ has a density.}

\noindent (1) {\bf (Random sinks)} \cite{LJ} {\it If
$\lambda_{\max}<0$, then with probability 1, $\mu_\omega$ is
supported on a finite set of points. }

\noindent (2) {\bf (Random strange attractors)} \cite{LY} {\it
If $\lambda_{\max}>0$, then with probability 1, $\mu_\omega$ is a
random SRB measure.}

\smallskip
SRB measures are special invariant measures that describe the
asymptotic dynamics of chaotic {\it dissipative} dynamical systems.
They live on unstable manifolds, which are families of curves,
surfaces etc. that wind around in a complicated way in the phase
space {\cite{er}}.

{\bf Consequences for reliability and unreliability:}  Since
reliability means the system's reaction to a stimulus is independent
of its initial state, it corresponds to a random fixed point, or 
$\lambda_{\max}<0$, as observed
in~\cite{reliable,reliable_and_amp_effects}.  Complicated sets
generated by unreliable dynamics have also been 
numerically observed before \cite{unreliable}.  
We point out here that the characterization
of unreliable dynamics by SRB measures has two important
consequences.  First, the very distinctive geometries of random
sinks and SRB measures provide a dichotomy between reliability and
unreliability that is easy to recognize in simulations (see Fig.~2).
Second, because the $\mu_\omega$ are singular (even though 
$\mu$ has a density), different initial conditions are attracted to a 
low-dimensional set that depends on the 
stimulus history.  This implies that even in unreliable dynamics, 
the responses are highly structured and far from uniformly 
distributed, a fact illustrated in  Fig.~\ref{f.rasters}(right).

Observe that since $\lmax$ is nonrandom, reliability is
independent of the realization $\omega$ once stimulus amplitude
is fixed.



\medskip
\centerline{\bf \small III. CONNECTION STRENGTHS,}
\centerline{\bf \small STIMULUS AMPLITUDE  AND RELIABILITY}

\medskip
There are many ways to quantify the degree of reliability.  In view
of the results in Sec.~II, we will use $\lambda_{\max}$.
Fig.~\ref{f.lyap} shows $\lmax$ as a function of feedback strength
$\afb$ and stimulus amplitude $\eps$.  Regions of reliability and
unreliability are clearly visible.  A closer examination of this
plot reveals many interesting phenomena that beg for explanations:

\begin{figure}[b!]
\includegraphics[bb=0.15in 0.15in 3.05in 3.05in]{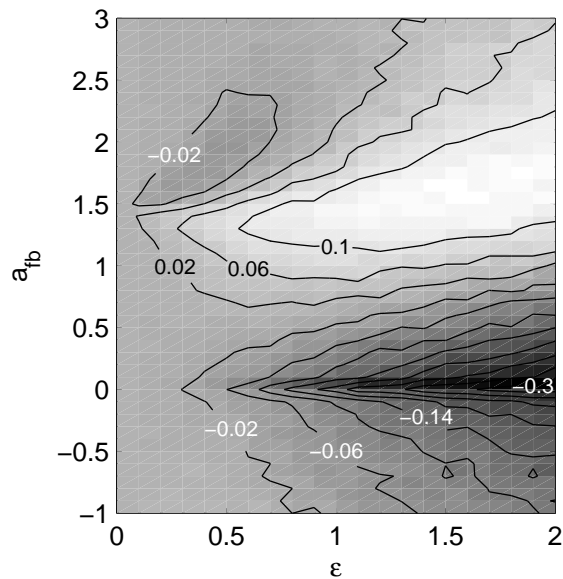}
\caption{Larger Lyapunov exponent $\la_{\max}$ as function
of $\afb$ and $\eps$ (with fixed $\aff=1$, $\omega_1=1$,
$\omega_2=1.1$).}
\label{f.lyap}
\end{figure}

(A) For a fixed value of $\afb$, how does increasing $\eps$
affect reliability?  Fig. 3 shows there is no simple
answer. When $\afb=0$, increasing $\eps$ makes $\lmax$ more
negative (a result consistent
with~{\cite{reliable_and_amp_effects}}), while at $\afb\approx
1.4$, $\lmax$ increases with $\eps$.  As if to confuse matters
further, for $\afb \in (1.5, 2.0)$, the system is reliable for
small $\eps$, unreliable as $\eps$ gets larger.

(B) Next we fix a value of $\eps$, say around $\eps=1.5$,
and ask how reliability properties vary with $\afb$.
Notice that $\lmax$ attains its minimum at $\eps=0$, i.e.
the purely feedforward network is the most reliable of all.
For relatively small values of $|\afb|$, negative feedback
is more reliable than positive feedback, and the system is
 unambiguously unreliable for a range of larger
positive $\afb$-values.

\begin{figure}
\begin{center}
\begin{tabular}{ccc}
\resizebox{1in}{!}{\includegraphics*[bb=0.03in 0.05in 1.97in 2.05in]{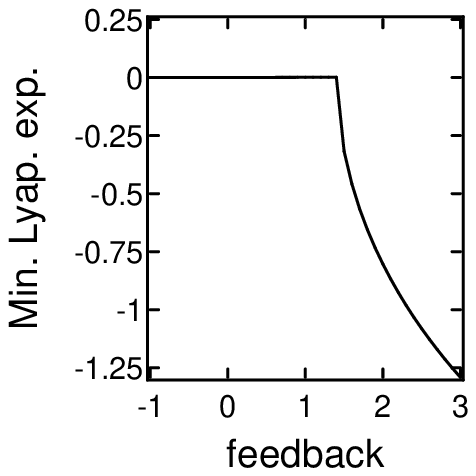}} &\hspace*{6pt}&
\resizebox{.7in}{!}{\includegraphics*[bb=12 27 133 241]{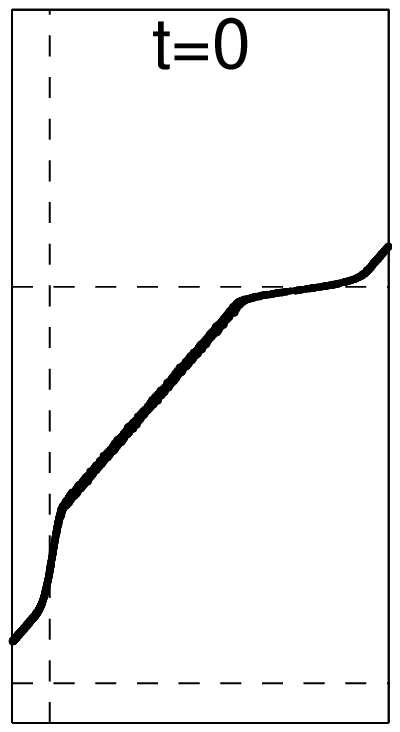}}
\resizebox{.7in}{!}{\includegraphics*[bb=12 27 133 241]{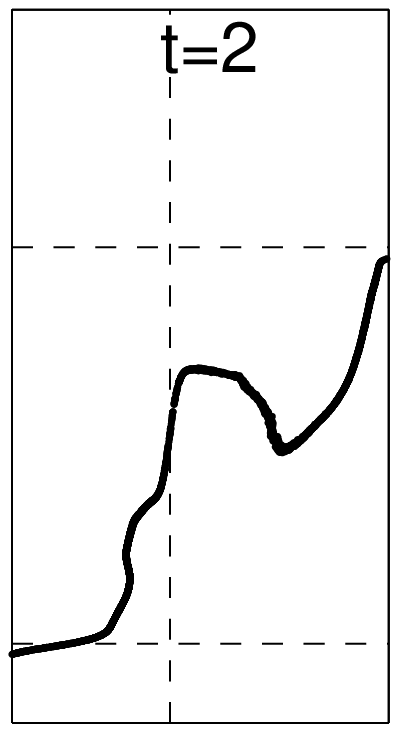}}
\resizebox{.7in}{!}{\includegraphics*[bb=12 27 133 241]{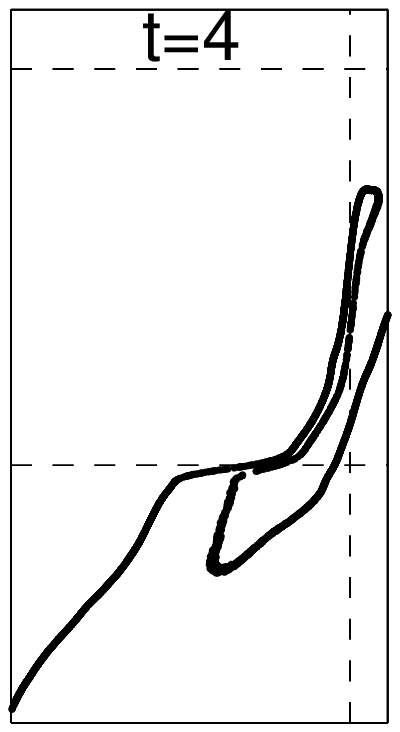}} \\
(a)&&(b)
\end{tabular}\vspace{-0.25in}
\end{center}
\caption{(a) Smaller Lyapunov exponent $\lmin$ for the
\textit{unforced} ($\eps=0$) network, as function of $\afb$. (b)
Folding action near limit cycle at $\afb=1.4, \eps=1.5$: at $t=0$,
the curve shown (with ends identified) is the limit cycle; the
second and third curves are images of the first, lifted to ${\mathbb
R}^2$. Dotted lines indicate the integer grid over the plotted
region. } \label{f.unforced}
\end{figure}

(C) We focus on the small-input region $\eps\approx 0$,
and observe the ``triple point'' near $\afb=1.4$, where a
small-amplitude stimulus has the largest impact on $\lmax$.
To explore why this configuration is especially vulnerable,
we study the zero-input system to look for clues near
$\afb=1.4$.
Fig.~\ref{f.unforced}(a) shows the smaller Lyapunov exponent
$\lmin$ for the system with $\eps=0$ as a function of $\afb$.
One sees that $\lmin=0$ up to about $\afb=1.4$, close to
where the triple point occurs. There it turns negative abruptly 
and begins to decrease \cite{lmin_footnote}.

Analogs of Figs.~\ref{f.lyap} and \ref{f.unforced}(a) (not shown)
for $\aff=1.5$ produce very similar results. The critical value of
$\afb$ differs, but there is a similar correspondence between a
triple point in the $\lmax$-picture and the transition in $\lmin$
for the zero-input system, suggesting that this is a general
phenomenon.


\medskip
\centerline{\bf \small IV. GEOMETRIC EXPLANATIONS}

\medskip

We review some well known facts from dynamical systems theory:
Chaos often results from stretch-and-fold actions; prototypical
examples are Smale's horseshoe and H\'enon's attractor.
Phase-space stretching is essential for the creation of positive
Lyapunov exponents, but it alone does not imply the existence of
strange attractors or SRB measures: Expansion is necessarily
compensated by contraction in some other directions or regions,
and contraction leads easily to stable equilibria or
sinks. There is tension between these forces; in general,
positive exponents prevail if there is a sustained expansion in
roughly identifiable directions. We now apply this thinking to
the context of Sec.~III.

{\it The 2-cell feedforward network is never unreliable:} In
Equation (\ref{e.tht}) with $\afb=0$ and any $\eps$, if
$\theta_1(0)=\bar \theta_1(0)$, then $\theta_1(t)=\bar \theta_1(t)$
for all $t$.  Geometrically, this says that the maps $F_{s,t;
\omega}$ in Sec.~II leave invariant the family of circles
$\{\theta_1=c\}$ in the torus ${\mathbb S}^1 \times {\mathbb S}^1$.
Since such a geometry precludes folding of any kind, no chaotic
behavior is possible.

{\it Role of unforced dynamics in unreliability:} The flow
generated by (\ref{e.tht}) with $\eps=0$ is a perturbation (due
to the coupling) of a linear flow with frequencies $\omega_1$
and $\omega_2$. Typically, the perturbed flow is either
quasi-periodic or possesses limit cycles. Fig. 4(a) tells us
that quasi-periodicity dominates for $\afb < 1.4$, beyond that
limit cycles prevail. In the region of $\afb > 1.4$ explored, we
find that there is a single cycle of period $\approx 1$ to which
all points in the phase space are drawn. Fig. 4(b) shows how
this cycle is folded by the maps $F_{0,t;\omega}$ for a randomly
chosen stimulus, setting the stage for the formation of strange
attractors.

It has been shown that limit cycles facilitate folding by
providing clear directions of potential expansion and
contraction.  Quasi-periodic dynamics do not provide such
preferred directions. In order for the folding to occur,
however, one must first ``break'' the cycle.

{\it Why the triple-point?}  Strongly attractive cycles are
harder to break and are likely to fold only when a stronger
stimulus is applied.  Weakly attractive cycles fold more
readily. Folding occurs, in fact, before a cycle is born, as
soon as phase points are attracted to a band.  These
observations explain why the onset of phase-locking in the
zero-input system coincides with $\lmax>0$ for weak $\eps$.  For
$\afb\gtrsim 1.5$, where the cycle is more robust ($\lmin$ is
more negative), the coupled system acts as a single --and hence
reliable-- phase oscillator, explaining the small ``island'' of
$\lmax<0$ seen there.

The ideas above come from the Wang-Young theory of periodically
kicked limit cycles, a rigorous theory explaining how small
folds created by kicks and amplified by shear lead to the
formation of strange attractors~\cite{shear}.  To determine to
what extent similar mechanisms are at work in Eq.~\ref{e.tht}
requires further analysis.

{\bf Conclusion.} We have demonstrated, via numerics and mathematical reasoning, 
the assertions stated in the Introduction.  Our main findings have implications for larger networks and for the encoding of stimuli.

\begin{small}
K.L. and E.S-B. hold NSF Math. Sci. Postdoctoral Fellowships and
E.S-B. a Burroughs-Wellcome Fund Career Award;
L-S.Y. is supported by a grant from the NSF.
\end{small}



\begin{thebibliography}{137}
\bibitem{initial_footnote} This is based on the presumption that
the system is intrinsically noisy.  To clarify the source of
unreliability in our model, the noise is turned off when $\eps
I(t)$ is presented; leaving this weak noise on has little
effect.
\bibitem{lmin_footnote} Closer examination reveals that $\lmin$
is, in fact, negative on some very tiny intervals for
$\afb<1.4$.
\end{thebibliography}
\end{document}